\documentclass[preprint]{aastex631}
 \usepackage[T1]{fontenc}      %ak
\usepackage[flushleft]{threeparttable}
\submitjournal{ApJ}

\shorttitle{On the Application of Differential Evolution to the Analysis of X-Ray Spectra}
\shortauthors{K{\k{e}}pa et al.}
\graphicspath{{./figures/}}
\begin{document}

\title{On the Application of Differential Evolution to the Analysis of X-Ray Spectra\footnote{Released on March, 1st, 2021}}

\email{ak@cbk.pan.wroc.pl}

\author[0000-0002-5299-5404]{Anna K{\k{e}}pa}
\affiliation{Space Research Centre (CBK PAN) \\
 Bartycka 18A, Warsaw \\
 Poland}

%collaboration{1}{(AAS Journals Data Scientists collaboration)}

%\author{Barbara Sylwester}
\author[0000-0001-8428-4626]{Barbara Sylwester}
\affiliation{Space Research Centre (CBK PAN) \\
 Bartycka 18A, Warsaw \\
 Poland}
                                                                                      
%collaboration{1}{(AAS Journals Data Scientists collaboration)}

\author[0000-0002-5006-5238]{Marek Siarkowski}
\affiliation{Space Research Centre (CBK PAN) \\
 Bartycka 18A, Warsaw \\
 Poland}
                                                                                       
%collaboration{1}{(AAS Journals Data Scientists collaboration)}
\author[0000-0002-8060-0043]{Janusz Sylwester}
\affiliation{Space Research Centre (CBK PAN) \\
 Bartycka 18A, Warsaw \\
 Poland}

%% Note that the \and command from previous versions of AASTeX is now
%% depreciated in this version as it is no longer necessary. AASTeX 
%% automatically takes care of all commas and "and"s between authors names.

%% AASTeX 6.31 has the new \collaboration and \nocollaboration commands to
%% provide the collaboration status of a group of authors. These commands 
%% can be used either before or after the list of corresponding authors. The
%% argument for \collaboration is the collaboration identifier. Authors are
%% encouraged to surround collaboration identifiers with ()s. The 
%% \nocollaboration command takes no argument and exists to indicate that
%% the nearby authors are not part of surrounding collaborations.

%% Mark off the abstract in the ``abstract'' environment. 
\begin{abstract}

Using methods of differential evolution (DE,) we determined the coronal elemental abundances and the differential emission measure (DEM) distributions for the plasma flaring on 2003 January 21. The analyses have been based on RESIK X-ray spectra. DE belongs to the family of evolutionary algorithms. DE is conceptually simple and easy to be implemented, so it has been applied to solve many problems in science and engineering. In this study we apply this method in a new context: simultaneous determination of plasma composition and DEM .
In order to increase the confidence in the results obtained using DE, we tested the use of its algorithms by comparing the DE synthesized spectra  with respective spectra observed by RESIK. Extensive discussion of the DE method used and the obtained physical characteristics of flaring plasma is presented.

\end{abstract}

%% Keywords should appear after the \end{abstract} command. 
%% The AAS Journals now uses Unified Astronomy Thesaurus concepts:
%% https://astrothesaurus.org
%% You will be asked to selected these concepts during the submission process
%% but this old "keyword" functionality is maintained in case authors want
%% to include these concepts in their preprints.

%% From the front matter, we move on to the body of the paper.
%% Sections are demarcated by \section and \subsection, respectively.
%% Observe the use of the LaTeX \label
%% command after the \subsection to give a symbolic KEY to the
%% subsection for cross-referencing in a \ref command.
%% You can use LaTeX's \ref and \label commands to keep track of
%% cross-references to sections, equations, tables, and figures.
%% That way, if you change the order of any elements, LaTeX will
%% automatically renumber them.
%%
%% We recommend that authors also use the natbib \citep
%% and \citet commands to identify citations.  The citations areIt is a great honour to be invited to became a member of
%% tied to the reference list via symbolic KEYs. The KEY corresponds
%% to the KEY in the \bibitem in the reference list below. 

\section{Introduction} \label{sec:intro}

The analysis of X-ray spectra is the powerful tool for plasma diagnostics. Physical parameters of the emitting coronal plasma, such as the values of temperature and emission measure, are not directly observed but can be determined from the modeling of the X-ray spectrum. For many sources the X-ray spectrum is not a unique function of a single temperature, but is described using a distribution of plasma with temperature, the so-called differential emission measure distribution (DEM, $\varphi(T)$). DEM connects the observed fluxes  with atomic data (characterizing the line/continuum emissivity) and the elemental abundance which is usually assumed from existing models. However, the differential emission measure thus has to be determined independently of the elemental abundances.  In this contribution we present a new approach in which DEM and elemental abundances are determined simultaneously.

The DEM distribution describes  the amount of the optically thin emitting plasma present in a given temperature interval ($dT$) and volume ($dV$) and is defined as $\varphi(T)=N_e^2\frac{dV}{dT}$, where $N_e$ is the plasma electron density. Knowledge of the DEM distributions is essential for the study of plasma heating, calculation of synthetic spectra, and prediction of the intensity of UV, EUV, and X-ray radiation of the Sun, as well as plasma radiation losses in the solar atmosphere.

The form of DEM can be determined by solving a set of the Fredholm's integral equations of the first case, in which the line/spectral band fluxes are defined as
\begin{equation}
F_i=A_i\int_{T} f_i(T,\lambda_i,N_e)\varphi(T) dT
\end{equation}
where $F_i$ is flux in appropriate spectral band/line $i$, $A_i$ represents the abundance of an element contributing to this flux, and $f_i(T,\lambda_i,N_e)$ is the known (atomic physics--CHIANTI in our case) emission function corresponding to the observed flux $F_i$.
The abundance values in Equation~1 are used in linear scale.  Further in the paper, the values of elemental abundances are given in the ''astronomical'' logarithmic scale (where log A\textsf{$_{H}$}=12) for easier comparison.

Equation  1  is an example of an ill-posed problem. The results of the  reconstruction of DEM distributions are  nonunique and subject to systematic errors \citep{Craig_1976Natur.264..340C}. However, the methods  available in the literature described recently, for example, by \cite{Aschwanden_2015SoPh..290.2733A}, \cite{Plowman_2013ApJ...771....2P}, \cite{Hannah_2012A&A...539A.146H}, \cite{Su_2018ApJ...856L..17S}, \cite{Kepa_2020SoPh..295...22K}, which have previously been tested on synthetic data, provide DEM solutions closely reproducing the observed input spectra as tested on synthetic models. The features present on DEM shape obtained from inversion of real data were interpreted in terms of separate magnetic loop volumes making up solar flare and active region plasmas. 

Since spectral line and continuum intensities depend directly on the absolute elemental abundance (see Equation~1), the results of any DEM inversion strongly depend on the assumed chemical composition of the plasma. Usually, the elemental composition assumed  in the inversion relies on earlier spectral observations or plasma composition determinations performed from in situ measurements (solar wind, solar energetic particles (SEPs)). The accuracy and applicability of these abundance determinations count on the calibration of the instruments that provided observations, physical models used in interpretation, and also the data statistics.  The uncertainties in elemental abundances strongly influence the results of the inversion process to an extent similar to the uncertainties inherent to spectral fluxes input to inversion. The assumptions used in theoretical derivation of line and continuum fluxes like ''Maxwellian plasma in ionization equilibrium'' may not be appropriate for some (transient) sources. Spectrally determined abundances depend, in addition,  on the accuracy of values for all the atomic parameters involved in spectra modeling like excitation rate coefficients, transition probabilities, etc. \citep{Doschek_2016ApJ...825...36D}.

The coronal elemental abundances (derived from solar wind particle measurements and UV and X-ray emission line observations) are known to differ from photospheric abundances according to the first ionization potential (FIP) of the element: the so-called ''FIP Effect''. Elements with FIP values below 10 eV (e.g., Si, Mg, Ca, Fe) are enriched by a factor of 2~$\div$~5 as compared with photospheric abundances \citep{Feldman_CHianti_1992ApJS...81..387F}. Elements with FIP values of more than 10 eV (e.g., C, N, O, Ne, Ar) have abundances that are equal to photospheric values or proxies (such as interplanetary plasma abundances). Earlier work by \cite{Meyer_1985ApJS...57..173M} and \cite{Feldman_CHianti_1992ApJS...81..387F} summarizes the FIP effect as it was known then for solar wind, SEPs, cosmic rays, and the solar corona itself. Observations of magnetically active stars having hot coronae \citep{Brinkman_2001A&A...365L.324B,Gudel_2001A&A...365L.344G} performed by the Chandra and XMM-Newton Reflection Grating Spectrometer  gave information about the so called  inverse FIP effect, in which  the coronal abundances of low-FIP elements are  decreased rather than enhanced relative to photospherical ones. The   abundance ratios of Ne (highest FIP) to Fe (low FIP) can be order of 10 compared to  the photospheric abundances of the Sun \citep{Gudel_2007LRSP....4....3G}. An inverse FIP effect was also found for the Sun, for example,  by \cite{Doschek_2015ApJ...808L...7D, Doschek_2016ApJ...825...36D} when the analyzing of the spectra of calcium and argon lines, and recently by \cite{To_2021ApJ...911...86T} from Hinode/Extreme-ultraviolet Imaging Spectrometer and by \cite{Katsuda_2020ApJ...891..126K} for silicon, calcium, sulfur, and argon abundances determined for four solar flares of GOES X                                                                class. In all cases, the inverse FIP effect was only found during flares in small areas of very complex active regions.

RESIK - REntgenovsky Spektrometr s Izognutymi Kristalami \citep{Sylwester_2005SoPh..226...45S} recorded solar spectra from 2001 to 2003  aboard the Russian Coronas-F spacecraft \citep{Kuznetsov_2014ASSL..400....1K} in four spectral bands covering the nominal range 3.3~\AA~--~ 6.1~\AA. In this spectral range the emission lines of hydrogen- and helium-like ions of such elements as Si, S, Cl, Ar, and K are observed. RESIK measurements constitute a unique database for analysis of DEM distributions of various solar structures (flares, active areas, quiet corona) and determination of the elemental abundances. The uniqueness of the RESIK data is related to very good intensity calibration of this instrument performed at Rutherford Appleton Laboratory and Mullard Space Science Laboratory,  a higher accuracy than for previous spectrometers. The accurate estimation of the level  of emission arising from  instrumental fluorescence, which is often a problem for X-ray spectrometers has been performed for RESIK.

DEM distributions calculated based on RESIK spectra are presented in a number of papers available in the literature  \citep{Sylwester_2005ESASP.600E.143S,Kepa_2006SoSyR..40..294K,Sylwester_2015IAUGA..2254762S, Chifor_2007A&A...462..323C,Dzifcakova_2008A&A...488..311D,Sylwester_Opt2015ApJ...805...49S}.
The abundances of the main elements contributing to these  spectra (potassium, argon, chlorine, sulfur, and silicon) were determined using two assumptions: the isothermal and
multithermal X-ray emission.  The results obtained  in the first approach are given in series of articles by  \cite{Sylwester_Ar_2010ApJ...720.1721S, Sylwester_K_2010ApJ...710..804S, Sylwester_Cl_2011ApJ...738...49S, Sylwester_S_2012ApJ...751..103S,  Sylwester_Si_2013SoPh..283..453S}, while those for the multithermal plasma assumption  are given in the papers by \cite{Sylwester_Opt_2014ApJ...787..122S, Sylwester_Opt2015ApJ...805...49S}. In this second case the abundances of Si, S, Ar, and K were optimized using the maximum-likelihood, Bayesian Withbroe-Sylwester (WS) inversion technique \citep{Sylwester_1980A&AS...40..335S}.  

In Table 1 we summarize values of abundances  determined using RESIK spectra and compare them with the values available  in the literature.

\begin{table*}[h!]

\caption{Element abundances solar photosphere and corona, and in meteorites (C1 chondrites).}
{\footnotesize
\begin{tabular}{lccccc}
\hline\hline
 Elements & K & Si & S & Cl &  Ar  \\
 FIP (eV) & 4.3 & 8.1 &  10.3 & 13.0 & 15.8  \\ \hline

\hline
\textit{Meteoritic abundances $^{*}$} & & & & & \\
\cline{1-1} \\
\cite{Lodders_2003ApJ...591.1220L} & -- & -- & 7.19$\pm$ 0.04  & -- & --  \\ 
\cite{Grevesse_2007SSRv..130..105G} & 5.06$\pm$0.05 & 7.51$\pm$0.02 & 7.16$\pm$0.04  & 5.23$\pm$0.06 &  --  \\ \hline
%\multicolumn{5}{c}{}{\textit{Photosperic abundances}} \\
% \cline{2-6}
\textit{Photosperic abundances} & & & & & \\
\cline{1-1} \\
\cite{Grevesse_2007SSRv..130..105G}  &  5.08$\pm$0.07  & 7.51$\pm$0.02 & 7.14$\pm$0.05  & 5.50$\pm$0.30 & 6.18$\pm$0.08 \\ 
\cite{Asplund_2009} & 5.03$\pm$0.09 & 7.51$\pm$0.03 & 7.12$\pm$0.03  & 5.50$\pm$0.30 & 6.40$\pm$0.13 \\ 
\cite{Caffau_2011SoPh..268..255C} & 5.11$\pm$0.09 & -- & 7.16$\pm$0.05  & -- & --  \\ \hline
 %\multicolumn{5}{c}{}{\textit{Coronal abundances}} \\
%  \cline{2-6}
\textit{Coronal abundances} & & & & & \\
\cline{1-1} \\
\cite{Feldman_CHianti_1992ApJS...81..387F} & 5.67$^{**}$ & 8.10 & 7.27 & -- & 6.58 \\
RESIK, isothermal assumption & 5.86 $\pm$0.23 & 7.93$\pm$ 0.21 (Si {\sc{xiv}} Ly$\beta$ line) & 7.16$\pm$ 0.17 & 5.75$\pm$ 0.26  & 6.44 $\pm$ 0.07 \\
                             &                & 7.89$\pm$ 0.13 (Si {\sc{xiii w3}} line)  & &   &  \\
RESIK, multithermal assumption & 5.73$\pm$0.19 & 7.53 $\pm$0.08 & 6.91$\pm$0.06 & --  &  6.47$\pm$0.08\\ \hline
\end{tabular}}
\begin{tablenotes}
\item ${*}$ The elemental abundance  values  are in the astronomical scale, in which hydrogen abundance is the reference element.   Meteoritic abundances typically are given in cosmochemical scale, in which the number of silicon atoms constitutes the reference, so these abundances were converted to the astronomical scale. \\
\item ${**}$\cite{Landi_2002ApJS..139..281L}  \\
\end{tablenotes}

\label{tab: label}
\end{table*}

Values  of calculated abundances  for argon and potassium  values in the case of isothermal and multithermal assumptions  based on RESIK observations are in agreement. For isothermal assumption the estimated sulfur abundance depends on the lines used in analysis and lies in the range from 7.13 to 7.24, with the most reliable value being 7.16 (estimated based  on the S~{xv w4} line). These values are very close to photospheric abundance estimates \citep{Asplund_2009, Caffau_2011SoPh..268..255C} and to quiet-Sun solar wind \citep{Reames_1999ApJ...518..473R}  and meteoritic abundances \citep{Lodders_2003ApJ...591.1220L}. The value of silicon abundance  depends on the spectral lines used for calculations. It differs by a factor of 2.6 and 2.4 from  the photospheric abundance \citep{Asplund_2009}, respectively.
The determined silicon  and sulfur abundances are higher than those obtained with the multithermal assumption. However, abundances obtained with the multitemperature approach should be preferred owing to their smaller uncertainties. In the case of sulfur, this lower abundance indicates that, in contrast to the view expressed by \cite{Feldman_CHianti_1992ApJS...81..387F}, sulfur behaves like a high-FIP element (the flare S abundance being even lower than in the photosphere). Chlorine abundance was determined with high error, which reflects a factor of two uncertainty in measured line fluxes.

In this contribution we propose a new approach for determination of elemental abundances from EUV and X-ray spectra. The new method solves the inversion simultaneously searching for the optimum combination of DEM distribution and the set of abundance values for elements important in the RESIK spectra formation. In the described approach a  differential evolution (DE) method  \citep{Storn_1997,
Price_Storn_2005_Springer} is used. DE is a powerful technique for difficult optimization problems and has been used in dozens of studies, e.g. for the identification of stellar flares \citep{Kellen_2019}, stellar classification \citep{Sanchez_9046495}, and the optimization of neutrino oscillation parameters \citep{Mustafa_2013}. We have used this method for the analysis of RESIK spectra.
The DE method has already been tested and used for determinations of  DEM distributions of flaring plasma \citep{Kepa_2016IAUS..320...86K, Kepa_2020SoPh..295...22K}. Here, this approach is applied to calculations of the elemental abundances for the first time. This novel analysis was performed for the example, well-observed RESIK flare that  occurred on 2003 January 21 ({\textsf{SOL2003-01-21T15:26}}, M1.9 GOES class).

This contribution is organized as follows: A description of the investigated flare and a short introduction to the DE method are given in Section 2. The algorithm test  and the results of analysis are shown in Section 3. The conclusions and discussion are provided in Section 4.  

\section{Observations and description of the method}

We studied the M1.9 GOES class flare that occurred on the east limb of the solar disc (S07E90) on 2003  January 21 and reached maximum in  X-ray at~$\sim$~15:26 UT ({\textsf{SOL2003-01-21T15:26}}). RESIK observations for this event are available from 14:56:35 UT to 18:56:34 UT with a few gaps when the \textit{Coronas-F} spacecraft crossed through the Van Allen  belts or the South Atlantic Anomaly. The RESIK spectrometer recorded  289 high-quality spectra with exposition from 4 to 905 s, and the total time for spectra collection was about 1.5 hr.
 
%\vspace{-1cm}
 \begin{figure}
 \centerline{\includegraphics[width=0.95\textwidth,clip=]{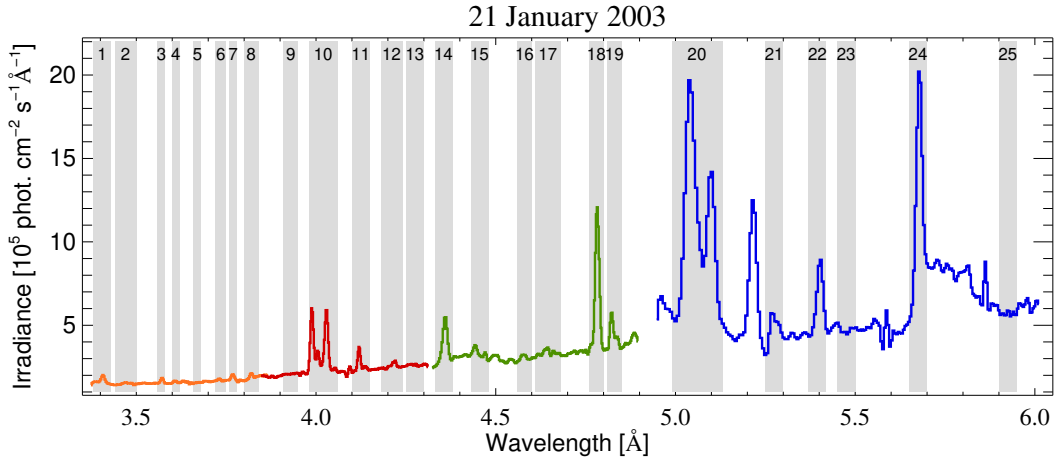}}
 \vspace{-14.5cm}
\caption{The spectrum of the 2003 January 21 flare integrated over 1.5 hr for four RESIK channels (1:~3.3~\AA~--~3.8~\AA, orange; 2:~3.8~\AA~--~4.3~\AA, red; 3: 4.3~\AA~--~4.9~\AA, green; 4: 5.0~\AA~--~6.1~\AA, blue).}
\label{fig:fig2}
\end{figure} 
\begin{table}[h!]
\caption{Wavelength Bands Used for Determinations of DEM and Elemental Abundances.}
\center
{
\begin{tabular}{c c c c @{} l} \hline \hline
{No.} & {Wavelength Range [\AA]} & {Main Line atop the Continuum} & {RESIK Channel} \\ \hline
{1} & {3.343~--~3.393} & {Ar~{\sc{xvii}} 1\textit{s}$^{2~1}$\textit{S}$_{0}$~--~1\textit{s}~3\textit{p}~$^1$\textit{P}$_{1}$} & {1} \\ %\hline
{2} & {3.403~--~3.463} & {Ar~{\sc{xvi}} 1\textit{s}$^{2}$~--~1\textit{s}~3\textit{p} sat.} & {1} \\ %\hline
{3} & {3.523~--~3.543} & {K~{\sc{xviii}} 1\textit{s}$^{2~1}$\textit{S}$_0$~--~1\textit{s}~2\textit{p}~$^1$\textit{P}$_1$} & {1} \\ %\hline
{4} & {3.563~--~3.583} & {K~{\sc{xviii}} 1\textit{s}$^{2~1}$\textit{S}$_0$~--~1\textit{s}~2\textit{s}~$^3$\textit{S}$_1$} & {1}\\  %\hline
{5} & {3.623~--~3.643} & {continuum} & {1} \\ %\hline
{6} & {3.683~--~3.713} & {S~{\sc{xvi}} 1\textit{s}$^{~2}$\textit{S}$_{1/2}$~--~5\textit{p}~$^2$\textit{P}$_{1/2,3/2}$ } & {1} \\ %\hline
{7} & {3.723~--~3.743} & {Ar~{\sc{xviii}} 1\textit{s}$^{2~1}$\textit{S}$_{1/2}$~--~2\textit{p}~$^2$\textit{P}$_{1/2,3/2}$} & {1} \\ %\hline
{8} & {3.763~--~3.803} & {S~{\sc{xvi}} 1\textit{s}$^{~2}$\textit{S}$_{1/2}$~--~4\textit{p}~$^2$\textit{P}$_{1/2,3/2}$} & {1} \\
\hline
{9} & {3.873~--~3.913} & {continuum} & {2} \\ %\hline
{10} & {3.943~--~4.023} & {Ar~{\sc{xvii}} triplet} & {2} \\ %\hline
{11} & {4.063~--~4.113} & {S~{\sc{xv}} 1\textit{s}$^{2~1}$\textit{S}$_{0}$~--~1\textit{s}~4\textit{p}~$^1$\textit{P}$_{1}$} & {2} \\ %\hline
{12} & {4.143~--~4.203} & {Cl~{\sc{xvii}} 1\textit{s}$^{~2}$\textit{S}$_{1/2}$~--~2\textit{p}~$^2$\textit{P}$_{1/2,3/2}$} & {2} \\ 
{13} & {4.213~--~4.263} & {continuum} & {2} \\
\hline
{14} & {4.278~--~4.328} & {S~{\sc{xv}} 1\textit{s}$^{2~1}$\textit{S}$_{0}$~--~1\textit{s}~3\textit{p}~$^1$\textit{P}$_{1}$} & {3} \\ %\hline
{15} & {4.378~--~4.428} & {S~{\sc{xiv}}~d~1s$^{2}$~2\textit{p}~$^2$\textit{P}$_{1/2,3/2}$~--~1\textit{s}~2\textit{p}~($^3$P)~3\textit{p}~$^2$\textit{D}$_{3/2,5/2}$} & {3}\\ % 
{16} & {4.508~--~4.548} & {continuum} & {3} \\ %\hline
{17} & {4.558~--~4.628} & {continuum} & {3} \\% \hline
{18} & {4.708~--~4.748} & {S~{\sc{xvi}} 1\textit{s}$^{~2}$\textit{S}$_{1/2}$~--~2\textit{p}~$^2$\textit{P}$_{1/2,3/2}$} & {3} \\ %\hline
{19} & {4.758~--~4.798} & {Si~{\sc{xiv}} 1\textit{s}$^{~2}$\textit{S}$_{1/2}$~--~6\textit{p}~$^2$\textit{P}$_{1/2,3/2}$} & {3} \\
\hline
{20} & {4.988~--~5.128} & {S~{\sc{xv}} triplet} & {4} \\ %\hline
{21} & {5.248~--~5.298} & {Si~{\sc{xiii}} 1\textit{s}$^{2~1}$\textit{S}$_{0}$~--~1\textit{s}~5\textit{p}~$^1$\textit{P}$_{1}$} & {4} \\ %\hline
{22} & {5.368~--~5.418} & {Si~{\sc{xiii}} 1\textit{s}$^{2~1}$S$_{0}$~--~1\textit{s}~4\textit{p}~$^1$\textit{P}$_{1}$} & {4} \\ %\hline
{23} & {5.448~--~5.498} & {continuum} & {4} \\ %\hline
{24} & {5.648~--~5.698} & {Si~{\sc{xiii}} 1\textit{s}$^{2~1}$\textit{S}$_{0}$~--~1\textit{s}~3\textit{p}~$^1$\textit{P}$_{1}$} & {4} \\% \hline
{25} & {5.898~--~5.948} & {continuum} & {4} \\ \hline
\end{tabular}
}
\end{table}
%\newline 
%\newline 

The average RESIK spectrum for the chosen flare is shown in Figure 1. The gray strips indicate 25 wavelength bands used for analysis. The information about emission lines observed by RESIK and about wavelength ranges selected is presented in Table 1. 
The different colors of the spectra represent four spectral channels (1:~3.3~\AA~--~3.8~\AA, orange; 2:~3.8~\AA~--~4.3~\AA, red; 3:~4.3~\AA~--~4.9~\AA, green; 4:~5.0~\AA~--~6.1~\AA, blue) in which RESIK spectra were recorded.  RESIK observations include the  1\textit{s}$^2$--1\textit{s} n\textit{p} and 1\textit{s}--n\textit{p} lines of He-like and H-like ions and  such elements as silicon, sulfur, argon, and potassium. For some flares the chlorine lines were visible. The lines of argon and sulfur were observed in the first and second channels. The potassium triplet (3.53~\AA) was observed in the first channel, and  the chlorine line was  noticed in the second channel \citep{Sylwester_Cl_2011ApJ...738...49S}.  In the third and fourth channels the lines of sulfur and silicon were the most pronounced. 

For the present analysis the fluxes containing lines and the  continuum were used, and the corresponding emission functions  have been determined using CHIANTI v8.0.7 \citep{Dere_1997A&AS..125..149D, Young_2019AAS...23431402Y}  and are described in more 
detail in Section 3. The spectral line Si~{\sc{xiv}}  1\textit{s}$^{~2}$\textit{S}$_{1/2}$~--~3\textit{p}~$^2$\textit{P}$_{1/2,3/2}$ observed around 5.2~\AA~(in channel 4) was excluded from the present analysis, as the intensity of this line is observed to be systematically too strong - for unknown reasons, probably related to partly non-thermal excitation of this line.

\vspace{1cm}
Here we introduce how the DE approach is adapted in order to be used for simultaneous determination of
DEM distributions and elemental abundances.

DE belongs to the class of evolutionary algorithms  designed to solve optimization problems with real-valued parameters. It is very powerful method for the black-box optimization, where the function to be optimized is very complex and it is difficult to find the appropriate  derivatives necessary to compute its extreme. DE is based on the  biological evolutionary processes: mutation, crossover, and selection.  

The values of three parameters are important to achieve the solution: a size of the population, and mutation ($F$), and crossover ($CR$) factors.

The most important characteristics of DE are the following:
\begin{itemize}
\item The mutation is provided by arithmetic combinations of individuals, not as the result of small perturbations to the genes as in genetic algorithm \citep{Feoktistov_2004}.
\item DE  works with  three equally large populations of individuals: the population of parents $S$, trial, and descendants.
\item The size of populations (a set of potential solution) - usually does not change during the evolution process.
\item The DE works with floating point numbers,  so the crossover/mutation is easier.
\end{itemize}
\begin{figure}[b!]
%\plotone{s_DE.png}
 \centerline{\includegraphics[width=0.5\textwidth,clip=]{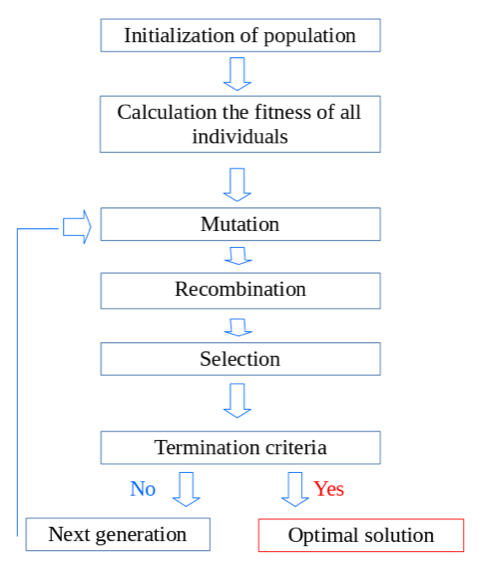}}
\caption{The classical DE algorithm.}
\label{fig:fig4}
\end{figure}
A typical working scheme for DE is presented in Figure 2. 
At the beginning  the population (\textit{S}) has to be initialized. In our case,  each  solution (individual) \textit{$s_i$} , where i~=~1,~2,~3 ... \textit{S}, represents the values of elemental abundances or DEM. The individuals consist of the genes.  This means that a few genes  are related to a chemical composition (one gene corresponds to the abundance for one element) and  a number of genes correspond to the values of DEM distribution. We use two models of the individuals with different representation of  DEM  distribution. In the first one (Figure 3a) genes represent  the abundances, temperatures, and DEM values. In the second model (Figure 3b) the genes represent abundances and DEM values only. In this approach  we set (define) values of temperatures at which DEM distribution will be determined.

\begin{figure}[h!]
%\plotone{s_DE.png}
 \centerline{\includegraphics[width=0.95\textwidth,clip=]{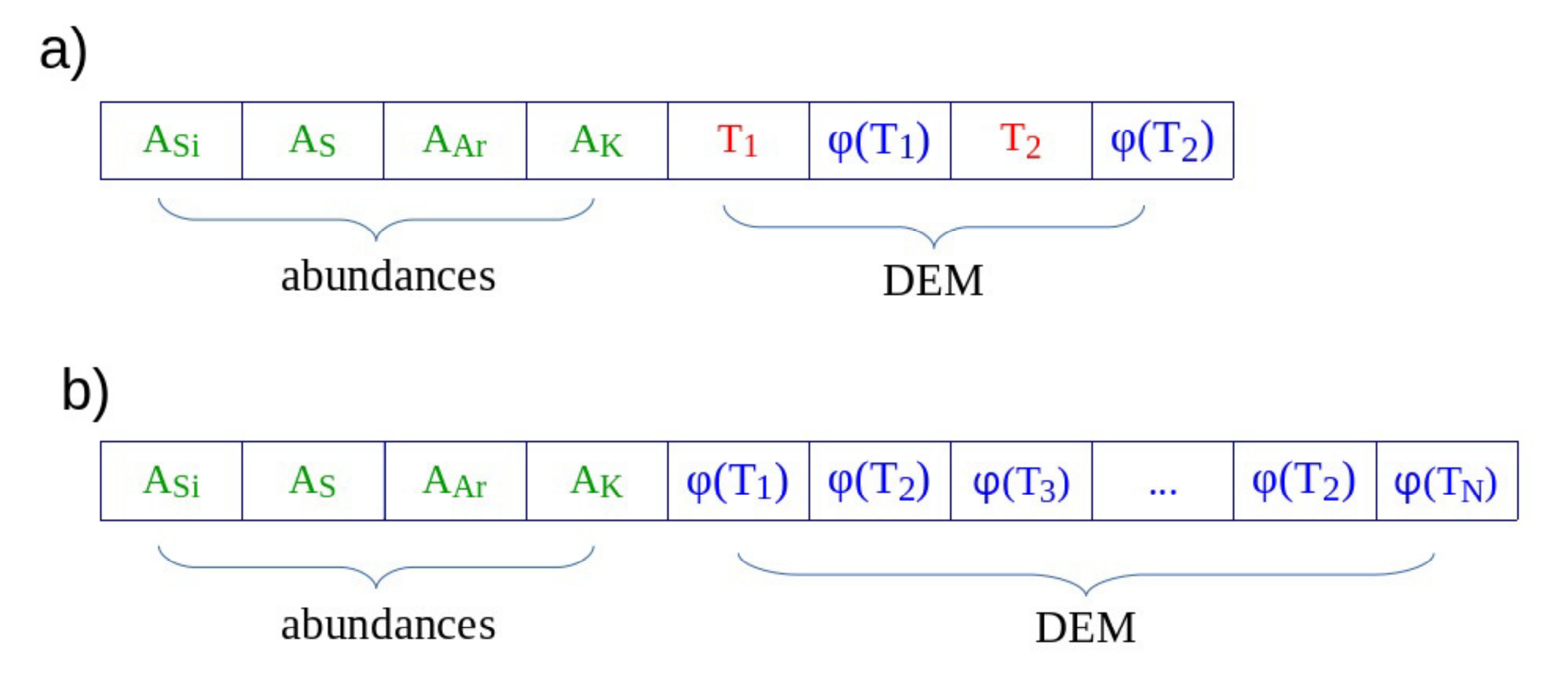}}
% \vspace{-10cm}
\caption{Schematic representation of two types of individuals which have been used in the analysis. One box in the diagram corresponds to one gene. A$_{Si}$, A$_{S}$,  A$_{Ar}$, and A$_{K}$ represents abundances of silicon, sulfur, argon, and potassium, respectively. $\varphi(T_n)$ represent DEM values for temperature $T_n$, where \textit{n}~=~1, 2, 3, ... \textit{N}.} 
\label{fig:fig3}
\end{figure}

For each parameter (gene; one box in Figure 3) the limiting  values have to be present and the initial population should cover the entire search space as much as possible.  In our specific case, each ($s_i$) solution (corresponding to one DEM distribution and one set of abundances of elements that dominate the spectra formation) is a randomly generated vector.

Next, the population is subjected to mutations.  In this process, for each individual in the population $S$ ($S$ is often referred to as the parent population)  the trial vector is created. It is important that  for each individual  exactly one individual of target population $U$ is created. In the basic version of mutation the target population is created  according to the following formula:
\begin{equation}
 \forall_{i, j} U_{i, j}=S_{i,j} + M\times(S_{r_1,j}-S_{r_2,j})
\end{equation}

where $S_{i, j}$ is the $j^{th}$ element of the $i^{th}$ individual, $M$ is the mutation factor, and $S_{r_1, j}, S_{r_2, j}$  are the $j^{th}$ elements of the other vectors $r_1$ and $r_2$. The $r_1$, $r_2$ indexes are randomly chosen and different from $i$ ($r_1 \ne  r_2$). The value of the $M$ factor affects the  exploration speed of the space solution; the greater value corresponds to the higher speed \citep{Georgioudakis_2020Frontiers}.

Besides the mutation scheme according to Equation~2, there are many other combinations  available in the literature \citep[e.g.][]{DAS20161}. However, in this paper we restricted ourselves to using the basic version of mutation. 

In order to increase the diversity of the perturbed parameter vectors, the population of target vectors ($U$) is subject to crossing. The crossover builds trial vectors $V_i$ through the recombination process from the elements of the parent vector, $S_i$, and the elements of the mutant vector, $U_i$ in the following way: 

\begin{equation}
\forall_{i, j} V_{i, j} =  \left\{ \begin{array}{ll}
U_{i, j} & \textrm{if} ~rand_j(0, 1) \leq CR,\\
S_{i, j} & \textrm{otherwise.}\\
\end{array} \right.
\end{equation}

%\forall_{i, j} V_{i, j} = 
%\begin{cases}
%\{U_{i, j} ~~~  if (rand_j(0, 1) \leq CR~~or~~j=j_{rand}},\newline
%S_{i, j} ~~~ otherwise.\\
%\end{cases}

Here the crossover probability, $CR \in $ [0, 1,] is user defined and $rand_j(0, 1)$ is a random number from the specified range.
 
To decide whether or not a given individual should become a member of next generation, 
the trial vector $V_i$ is compared to its direct parent $S_i$. If $V_{i}$ has a lower or equal objective function value than that of its target vector, $S_{i}$, it replaces the target vector in the next generation:

\begin{equation}
 \forall_{i} S_{i}^{k+1} = \left\{ \begin{array}{ll}
  V_{i}^{k} ~~~ & \textrm{if}~~~  f(V_i^k) \geq f(S_i^k), \\
 \newline
 S_{i}^{k} ~~~ & \textrm{otherwise.}\\
\end{array} \right.
\end{equation}

Here $S_i^{k+1}$ is the $i^{th}$  individual in the $k+1$ generation and $f$ is the objective function.

Our objective function is given as the ${\chi}^2$ value, defined as 
\begin{equation}
 \textit{f}={\chi}^2=\sum_{i=1}^{L}\frac{(F_i-F_{ic})^2}{\delta_i^2}
\end{equation}
 where:   $F_i$ is the observed flux in the $i^{th}$ spectral band from Table~2, $F_{ic}$ is the  flux in the $i^{th}$ line/spectral band calculated using Eq.~1 (based on $\varphi(T)$ distribution and abundance values taken from the individual) and $\delta_i$ is interpreted as an error consisting of the uncertainties in the measurements, and $L$ is the number of lines (or bands) used in the analysis ($L$=25 in our case). 
Our aim is to minimize the ${\chi}^2$ value.
The evolutionary process is repeated until the convergence criterion is fulfilled.
 
The choice of the optimal value  of parameters for DE method is a challenging task \citep{Chakraborty_Storn2008}. In the literature there exist many proposals of self-adaptive methods that  select the values
for M and CR \citep{Milani_2020}.  The parameters used in our study were adopted based on the tests performed on synthetic models. In our calculations we set mutation and crossover factors to be 0.1. The examples of the tests  using DE  for  simultaneous calculation of DEM distributions and plasma composition  are presented in the next section.
\\
\\

\section{Tests of spectra reconstructions}

DEM distribution  is derived  by deconvolving $\varphi(T)$ from the fluxes observed in selected spectral ranges (Equation 1) and known from theory theoretical emission functions $f_i(T,\lambda_i,N_e)$. In order to
be able to interpret the results of DE inversion, we performed numerous tests of spectra reconstruction for spectra generated for model sources. Testing provides insight into the reliability of the differential evolution approach for simultaneous determination of element abundances and differential emission measure distribution.

%\vspace{-4cm}
\begin{figure}[hb!]
%\plotone{s_DE.png}
%\hspace{1cm}
 \centerline{\includegraphics[width=15cm,clip=]{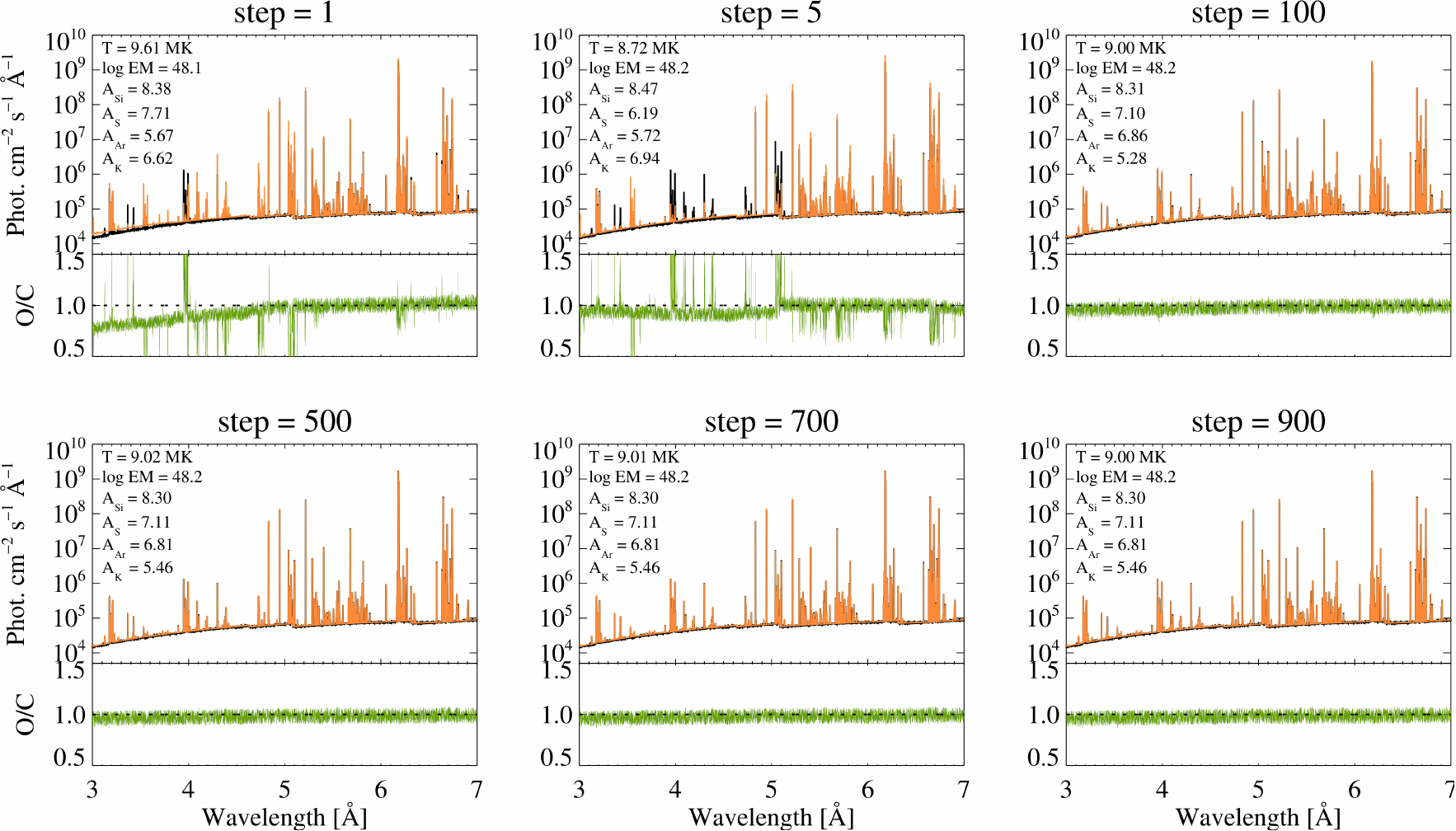}}
% \vspace{-5cm}
 \caption{Example of spectrum reconstructions using the one-temperature model of plasma in six selected iteration steps of DE method. The spectrum adopted as observation is in black, and in orange the best-fit calculated spectra for each given iteration step are presented. In green  the ratio of ''observed'' to ''the best'' spectrum is represented.  The spectrum assumed as ''observed'' has been calculated for the isothermal plasma at temperature T~=~9~MK  and emission measure EM~=~2$\times$10$^{48}$~cm$^{-3}$ for A$_{Si}$~=~8.3, A$_S$~=~7.12, A$_{K}$~=~5.41, A$_{Ar}$~=~6.82, respectively. Inserted in the left upper corner are the values of the temperature, emission measure, and abundances of silicon, sulfur, argon, and potassium obtained in the presented iteration step.}
\label{fig:fig4}
\end{figure}
\begin{figure}[h!]
%\plotone{s_DE.png}
%\hspace{1cm}
%\vspace{-12cm}
 \centerline{\includegraphics[width=15cm,clip=]{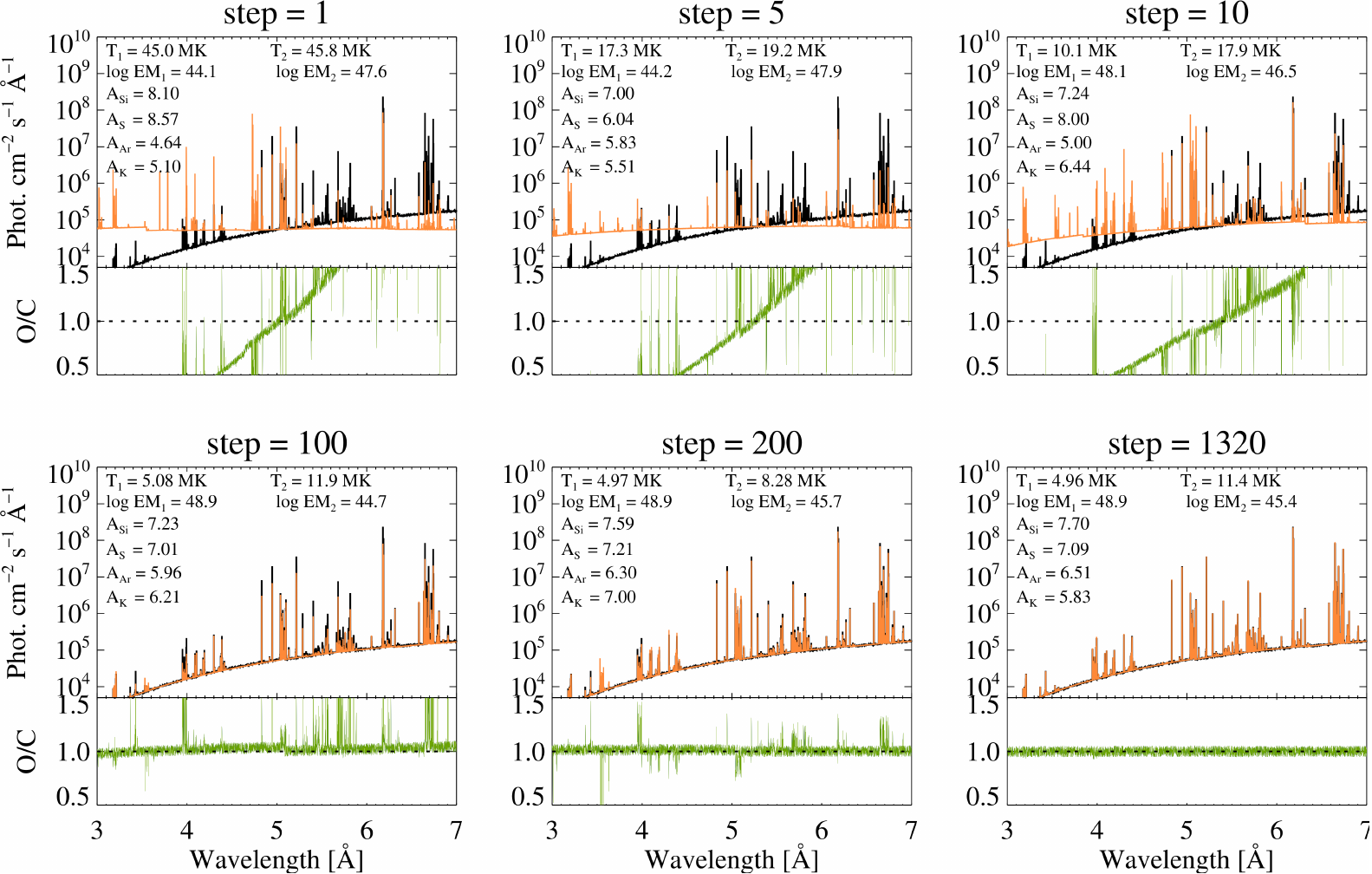}}
 %\vspace{-1.5cm}
 \caption{Example of spectrum reconstruction using the two-temperature model of plasma in six selected iteration steps of the DE method. The spectrum adopted as observations is in black, and in orange best-fit calculated spectrum for each given iteration step is presented. Green color represents the ratio of ''observed'' to ''the best'' spectra. We assumed  different amounts of plasma EM$_1$~=~1$\times$~10$^{49}$~cm$^{-3}$ and EM$_2$~=~2$\times$10$^{45}$~cm$^{-3}$ at T$_1$~=~5~MK and  T$_2$=12~MK, respectively and elemental abundances of silicon, sulfur, argonm and potassium as A$_{Si}$~=~7.70, A$_S$~=~7.10, A$_{K}$~=~5.91, and A$_{Ar}$~=~6.50. In the left upper corner are the values of the temperature, emission measure, and abundances of silicon, sulfur, argon, and potassium obtained in presented iteration step.}
\label{fig:fig4}
\end{figure}
We have used two types of source models, the isothermal and the two-temperature, and several sets of elemental abundances.  In this respect we assumed the ''coronal extended'' \citep[as in CHIANTI, after][]{Feldman_CHianti_1992ApJS...81..387F} for  most of the elements. The abundances for silicon, sulfur, argon, and potassium were drawn from the
values within specified limits in various exercises. The generated X-ray spectra were perturbed within limits given by errors coming from expected count statistics. Next, these spectra have been  treated as observations, which we should reconstruct using the DE method.
In the analysis we used the fluxes and corresponding (integral kernel) emission functions determined for 25 spectral bands presented in Table~2 that contain the contribution of emission in line and continuum. The emission functions have been calculated using CHIANTI package ver. 8.0.7 available in SolarSoft for ionization equilibrium based on \cite{Bryans_2009ApJ...691.1540B}  and unit elemental abundances.  Thanks to this, the change in the elemental abundances during each step of iteration of the DE method was associated with the multiplication of emission functions by the appropriate values only.
The ranges of values in which parameters could change were assumed to be  T~=~1--30~MK, log EM~=~ 40--50,  A$_{Si}$~=~7--9, A$_S$~=~6--8, A$_{K}$~=~5--7, and A$_{Ar}$~=~5--8. Abundances were expressed logarithmically on a scale with A$_H$ = 12. We started each test by generating a ''population'' of 500 trial solutions (S = 500), where each of the individuals has been defined according to the scheme presented in Figure 3a. The values for genes were random and placed within assumed ranges.

In Figure 4 we  illustrate the process of convergence between reconstructed and generated synthetic spectra at six indicated iteration steps during DE progress. For each iteration step we present the ''best-fit'' calculated spectrum and plot it over the synthetic ''observed'' one. The spectrum assumed as ''observed'' has been calculated for the isothermal plasma at temperature T~=~9~MK  and emission measure EM~=~2$\times$10$^{48}$~cm$^{-3}$. We assumed the abundances of silicon, sulfur, potassium, and argon  to be A$_{Si}$~=~8.3, A$_S$~=~7.12, A$_{K}$~=~5.41, and A$_{Ar}$~=~6.82, respectively.  The generated input synthetic spectrum was randomly perturbed by $\sim$10~$\%$ fluctuation in spectral irradiance. 

The analysis of the presented  test and many other tests made with the isothermal plasmas reveals that the DE method satisfactorily reconstructs the assumed plasma parameters after 500 steps of iteration. The obtained results are stable and do not change if the iterations continue (see the results presented for 700$^{th}$ and 900$^{th}$ steps). A similar pattern of convergence is observed for different sets of temperature and emission measure pairs as well as for different sets of abundances.

%\vspace{2cm}
In Figure~5 we present the sequence of spectra reconstruction obtained when the input synthetic spectrum was calculated for the
two-temperature model of the source. In the model we assumed a very small amount of high-temperature plasma EM$_2$~=~2$\times$10$^{45}$~cm$^{-3}$ at T$_2$=12~MK) ''contaminating'' the main cooler component of (EM$_1$~=~1$\times$~10$^{49}$~cm$^{-3}$ at T$_1$~=~5~MK. The elemental abundances of silicon, sulfur, argon, and potassium were taken to be A$_{Si}$~=~7.70, A$_S$~=~7.10, A$_{K}$~=~5.91, and A$_{Ar}$~=~6.50. In this case our population consists of  500 individuals (possible solutions) and the calculation process was interrupted after 5000 steps. 

\begin{figure}[th!]
%\plotone{s_DE.png}
%\hspace{1cm}
 \centerline{\includegraphics[width=21cm,clip=]{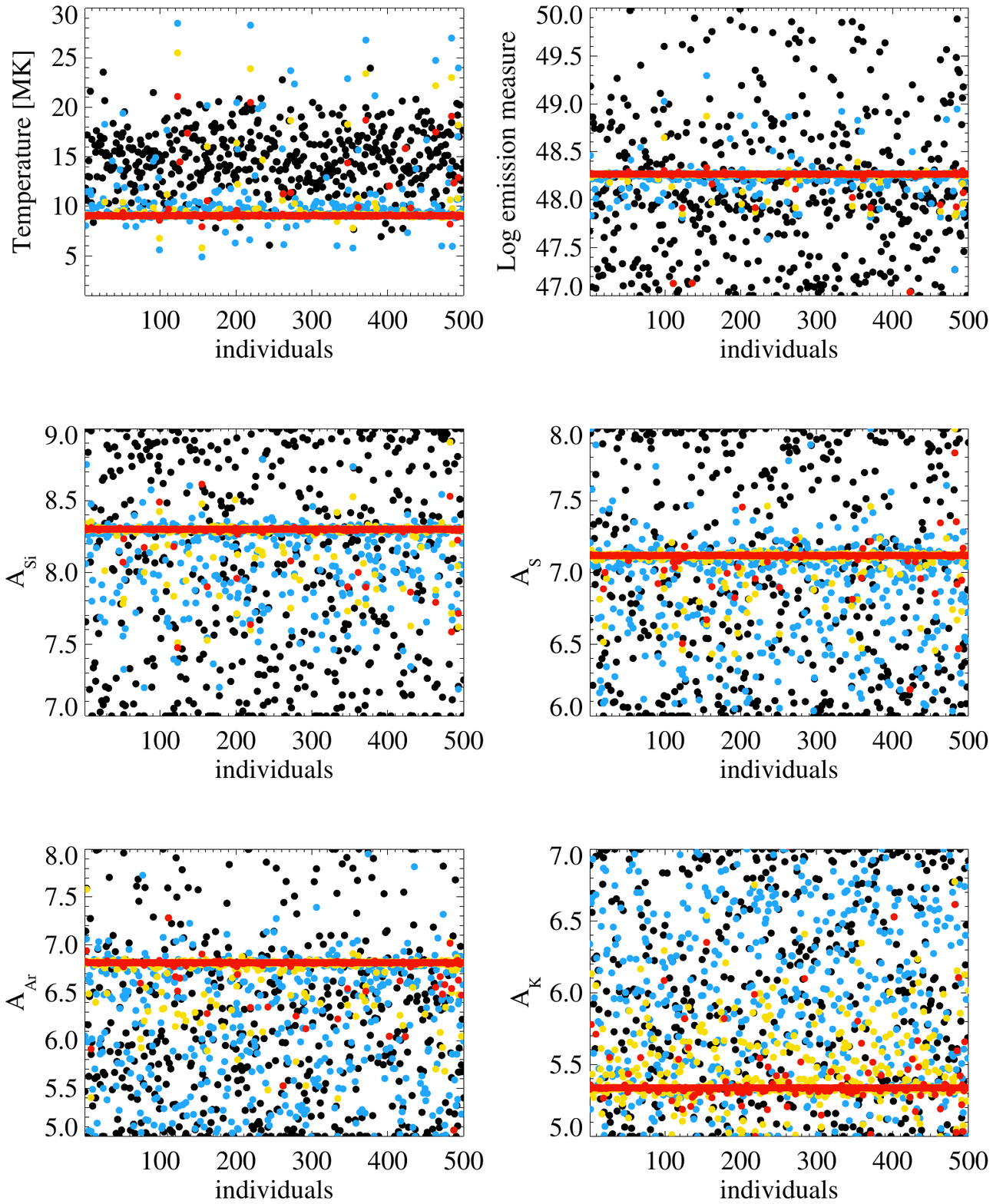}}
 \vspace{-6cm}
\caption{The values of temperature, emission measure, and silicon, sulfur, potassium, and argon abundances in the whole population of solutions in four selected iterations steps, steps 5 (in black), 200 (in blue), 500 (in yellow), and 1000 (in red) are  presented. The illustration is for the isothermal case (Figure 4).}
\label{fig:fig4}
\end{figure}

\begin{figure}[th!]
%\plotone{s_DE.png}
%\hspace{1cm}
 \centerline{\includegraphics[width=21cm,clip=]{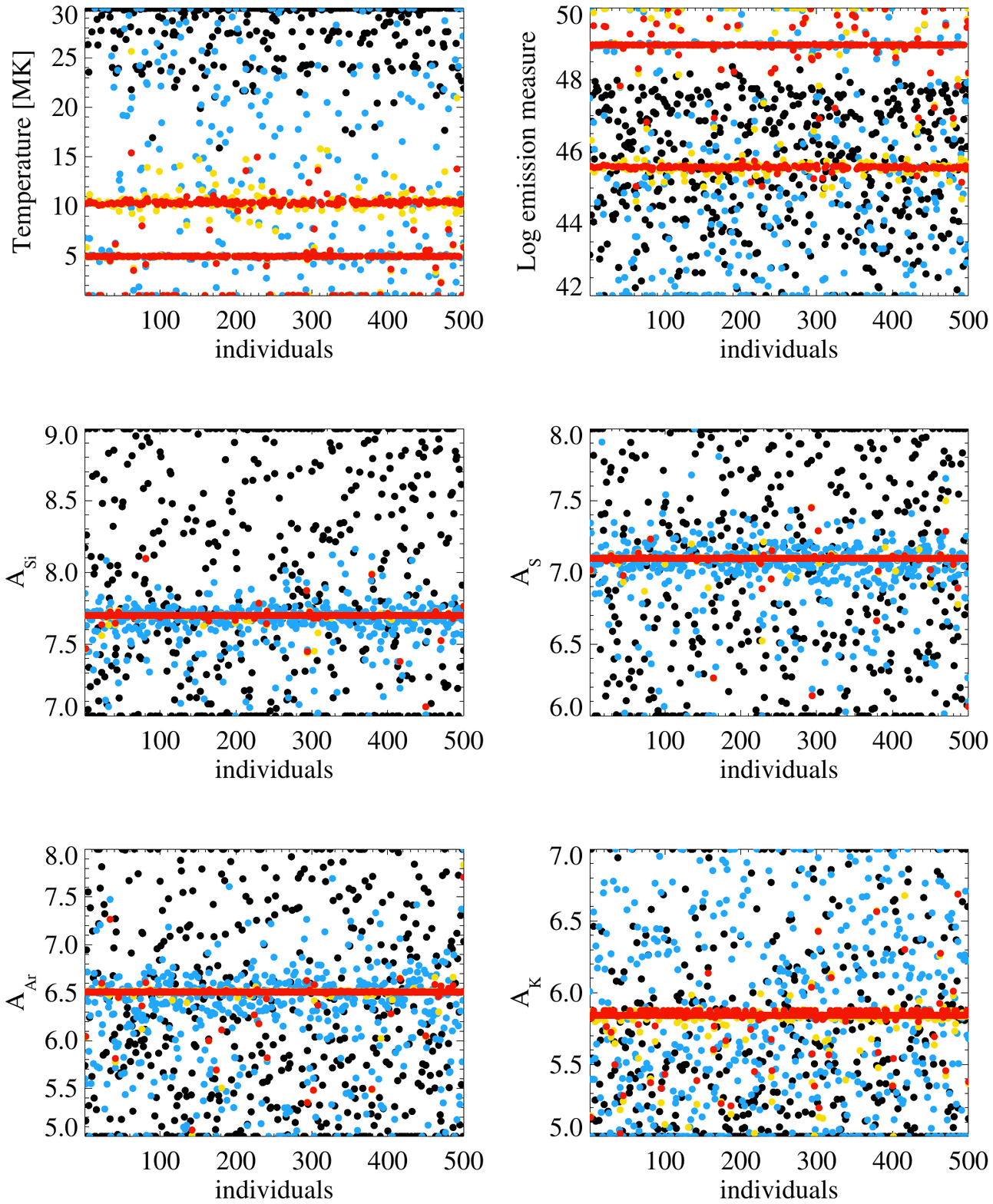}}
 \vspace{-6cm}
\caption{The values of temperature, emission measure, and silicon, sulfur, potassium, and argon abundances in the whole population of solutions in four selected iterations steps, steps 5 (in black), 500 (in blue), 3000 (in yellow), and 5000 (in red) are  presented. The presented results correspond to the two-temperature model from Figure 5.}
\label{fig:fig4}
\end{figure}

The results of the tests are similar to those obtained from the reconstruction of spectra calculated for the isothermal model of plasma. The spectrum has been reconstructed properly, and the obtained solution is in agreement with the assumed input parameters of the model. However, achievement of a good solution needs more iteration steps. In the presented case it was over  1300.  The calculations take few minutes on a personal computer.

In Figure 4 and Figure 5 we show the best solutions obtained in  selected iteration steps. However, each individual in the population is changed by mutations, crossover, and selection during the iteration. In Figure 6 and 7 we show how
the solutions differentiate within the  whole population  at four selected iteration steps  for isothermal and two-temperature models of plasma, respectively. Each parameter (temperature, emission measure, and silicon, sulfur, argon, and potassium abundances values) is presented in a the separate panel. Different colours represents different iteration steps.

The  differentiation of the whole population during reconstruction of the spectrum obtained with the isothermal model of plasma (from Figure 4) is presented (in Figure 6) for the steps  5 - in black, 200 - in blue, 500 in yellow, and 1000 in red.  As can be seen after 1000 steps, the entire population becomes almost homogeneous (all 500 solutions are similar), but 200 iterations are enough to find a proper solution. It is observed that the temperature structure (i.e. T and EM values) converges the fastest.  The potassium abundance value is established as the last parameter owing to a very weak A$_{K}$ and a weak contribution of
potassium lines to the single-band flux (the H-like K~{{\sc{xviii}}} triplet at  ~3.53~--~3.57~\AA~ contributing to bands No. 3 and 4). Potassium lines extend only slightly above the continuum, contrary to the other bands, where strong lines of Ar, S, and Si dominate the fluxes in bands.

The similar plots that present the results of spectrum reconstruction for a two-temperature model from Figure 5 are shown in Figure 7. Here, the state of "population differentiation" of solutions is presented as obtained in iteration steps 5, 500, 3000, and 5000.

One can see that for the two-temperature source the population becomes homogeneous after a longer evolutionary process than for the isothermal model. It is clearly seen that even in the last iteration shown (5000) a few percent of solutions still are different from assumed input parameters. This reflects the more complicated two-temperature "nature" of the source. The population differentiation is slightly dependent, however on the random draw of the initial population for the DE evolution. For various populations a satisfactory solution is obtained in different iteration steps. The process of evolution was stopped after 5000 generations because the convergence became very slow. The minimum $\chi^2$ in each case was below 1.
\begin{figure}[h!]
  \centerline{\ \includegraphics[width=16cm]{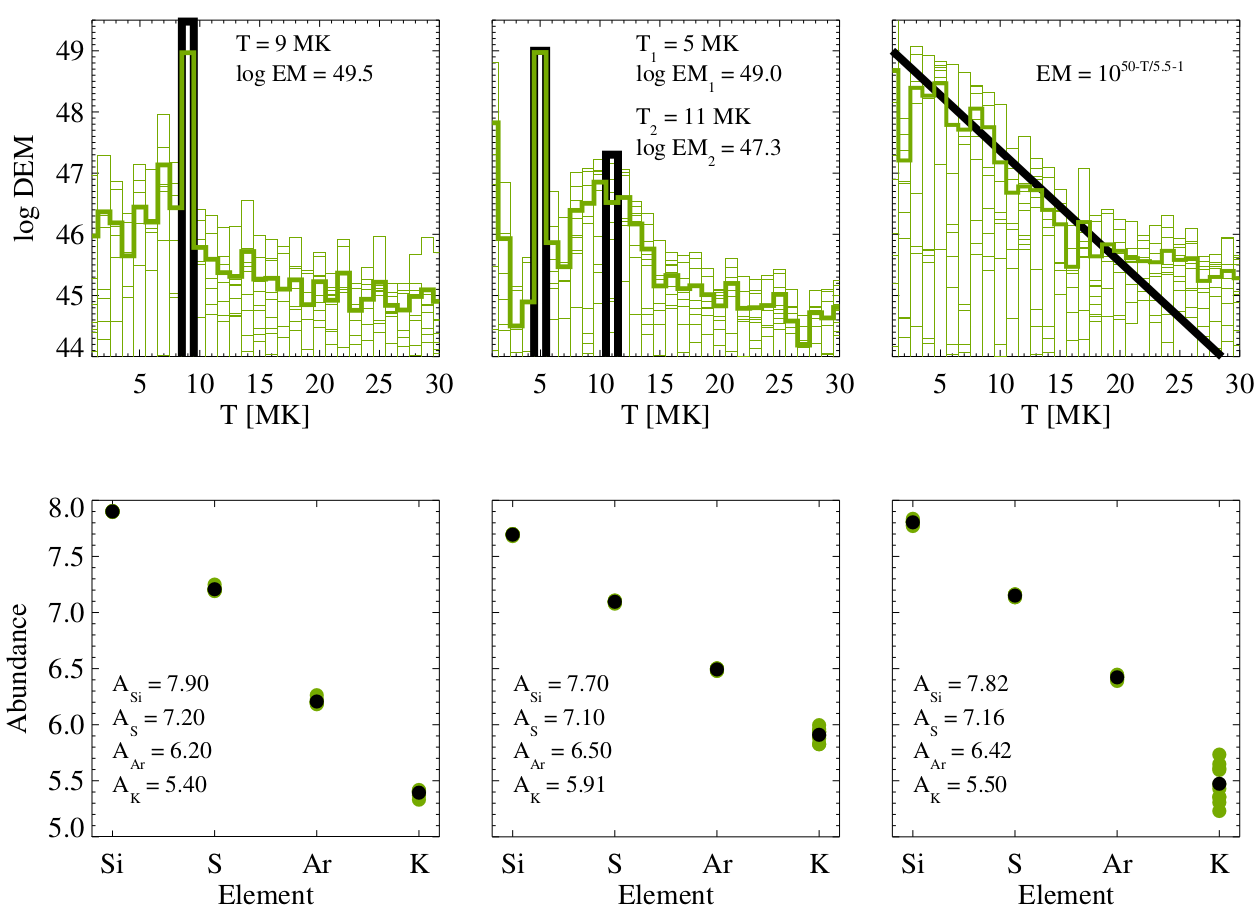}}
  \caption{Reconstruction of isothermal, two-temperature, and power law models of plasma  when DEM is described in 30 temperatures (scheme from Figure 3b).   Each column (upper plus lower panel) corresponds to one assumed model. The parameters of the models are given in the panels.  Black color is related to assumed models of plasma; green lines correspond to the calculated models. Top panels: in light green  lines  we show the results of 10 independent realizations of assumed models reconstruction  when the fluxes were perturbed  by random error 10 \% of the value.  The average models are presented in dark green. Bottom panels:  the comparison of the assumed and obtained abundances of silicon, sulfur, argon, and potassium.}
\label{fig:fig}
\end{figure}

We checked also the possibility of simultaneous calculation of elemental abundances and DEM distribution for the multitemperature approach. We have calculated DEM for 30 temperatures (30T) with $\Delta$T = 1~[MK] in the range 1 to 30~[MK] (using the DE method). In this case, the population consisted of individuals created according to the scheme from Figure 3b.  The temperatures in the DE method were predefined, and the sought variables were the abundances and emission measures in these 30 temperatures. The obtained results of reconstruction for three different models of plasma distribution are given  in Figure 8. 

Based on the results of the assumed model  reconstructions  using the multithermal approach, we can conclude that DE constitutes a viable alternative method not only for determination of DEM distributions but also for inferring values of elemental abundances from well-calibrated X-ray spectra. All unknowns (DEM and composition) are being reconstructed simultaneously, but at various paces, reproducing
 satisfactorily not only the shape of the DEM but also the set of input elemental abundances.  The method correctly reproduces the saturation values of silicon, sulfur, and argon. However, the value of the abundance of potassium is burdened with a very large error owning to the sensitivity limit of this method  (in the case of multitemperature plasma).

\section{DEM Distributions and Abundances for 2003 January 21 flare}

After testing on the synthetic models, the DE method was used  for the study of the flare that occurred on 2003 January 21 with maximum in X-ray at 15:26~UT. In Figure~9, we show  the temporal evolution of the soft X-ray fluxes in GOES bands 1~--~8~\AA~ and 0.5~--~4~\AA. All time intervals when RESIK operated and recorded spectra are marked by gray strips.  The temporal evolutions of isothermal temperature and emission measure as calculated based on GOES data are presented too. The values of temperature and emission measure were derived using the filter-ratio technique (with the flare background level subtracted) using the  standard routine \textsf{goes.pro}  from  SolarSoft package. The determined  maximum values of temperature and emission measure were 14.9~MK (at 15:04 UT) and  1.37$\times$10$^{49}$~cm$^{-3}$ (about 15:29 UT) respectively.

\begin{figure}[h!]
%\plotone{s_DE.png}
 \centerline{\includegraphics[width=0.8\textwidth,clip=]{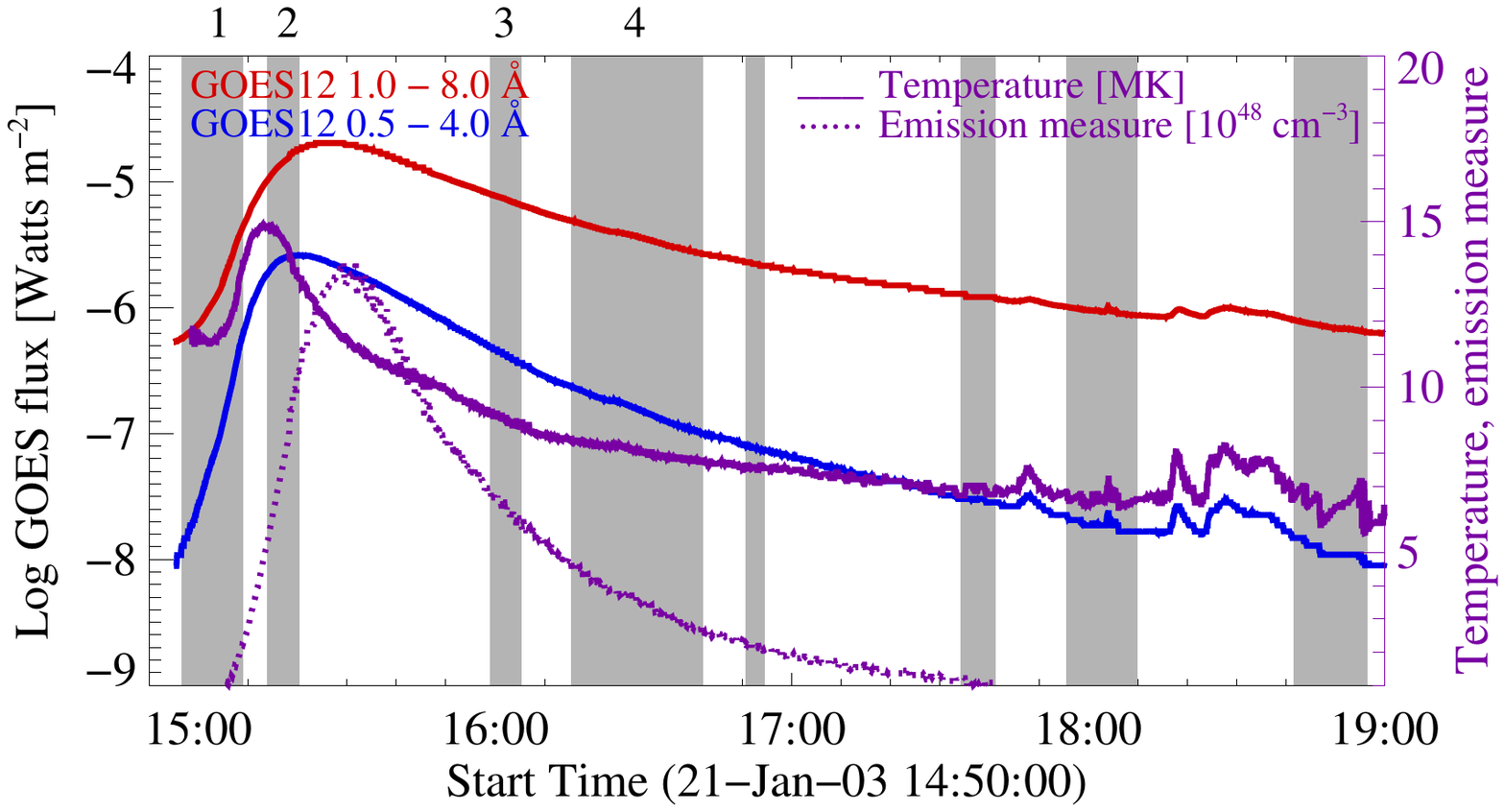}}
\vspace{-11.5cm}
\caption{GOES (in red and blue) light curves for the \textsf{SOL2003-01-21T15:26} flare. The  gray strips indicate time intervals when RESIK spectra are available. The purple lines represent the temporal variations of temperature and emission measure calculated based on GOES fluxes using the isothermal approximation. The time intervals analyzed using the DE algorithm are marked by the numbers. The left-hand y-axis should be referred to the GOES fluxes. The right-hand y-axis should be referred to the temperature in MK and emission measure in units of 10$^{48}$~cm$^{-3}$.} 
\label{fig:fig1}
\end{figure}

\begin{figure}[th!]
\vspace{-5cm}
 \centerline{\includegraphics[width=19cm,clip=]{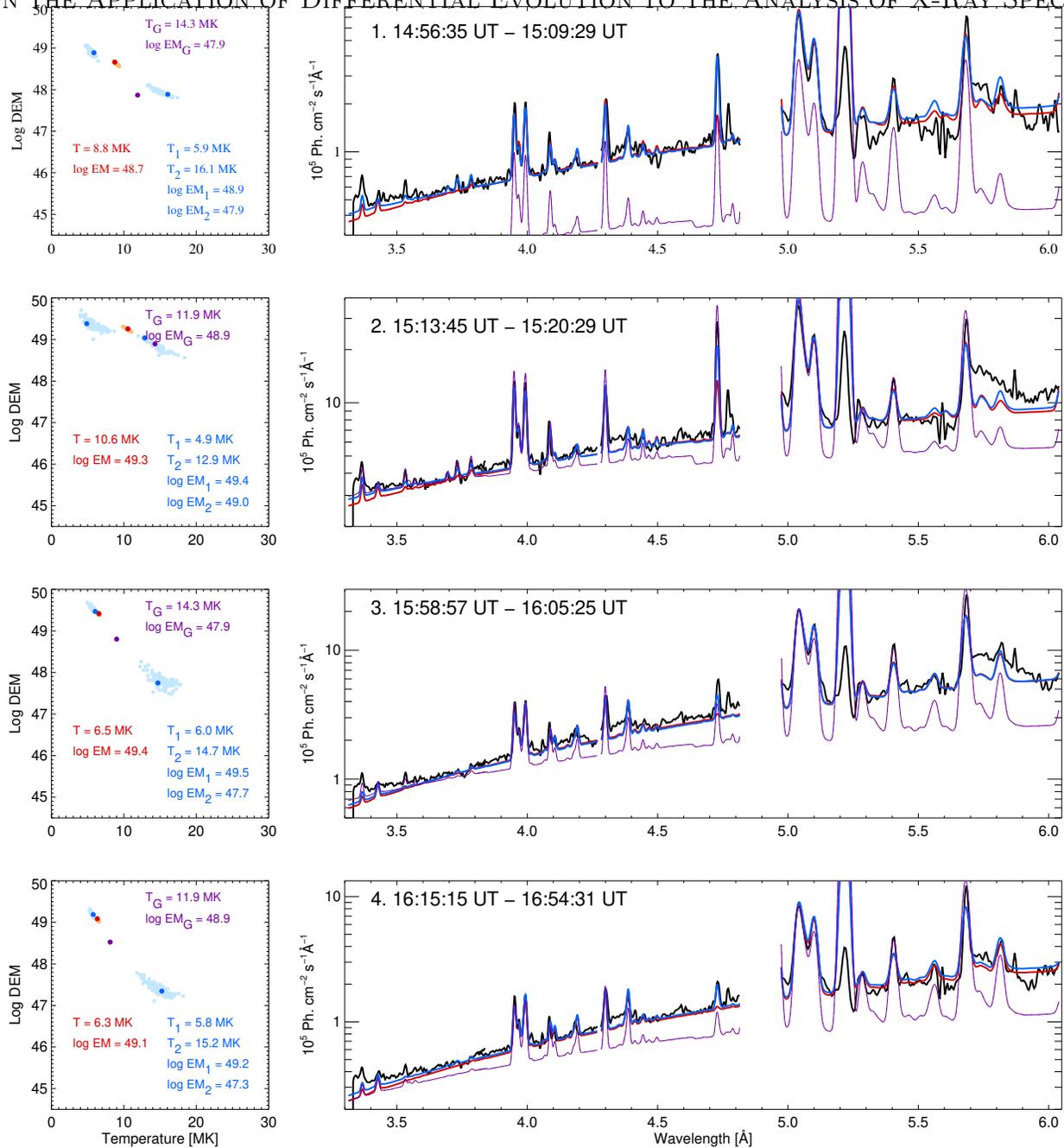}}
\vspace{-2cm}
\caption{The sequence of derived DEM distributions (left panel) and RESIK measured and calculated spectra (right) taken
at four selected time intervals. Each row corresponds to one time interval. Colors red and blue are related to
isothermal and two-temperature models of plasma, respectively. For each time interval for DE 100 differential
emission measure distributions that correspond to 100 independent generations of the population are presented. Additionally, the observed fluxes were randomly perturbed (10\% error, independently for each population). The light-red and blue colors show solutions obtained based on the perturbed fluxes. The dots represent the results for undisturbed fluxes. Dark-purple dots represent corresponding temperature and emission measure from GOES. In the right panel the RESIK-observed spectra are presented in black. Synthetic spectra (in red and blue for isothermal and two-temperature model of plasma, respectively) were calculated based on unperturbed fluxes. Dark-purple color represents temperature, emission measure, and the synthetic spectra obtained from analysis of GOES fluxes (see text for details).}
\label{fig:fig4}
\end{figure}

\begin{figure}[th!]
 \centerline{\includegraphics[width=19cm,clip=]{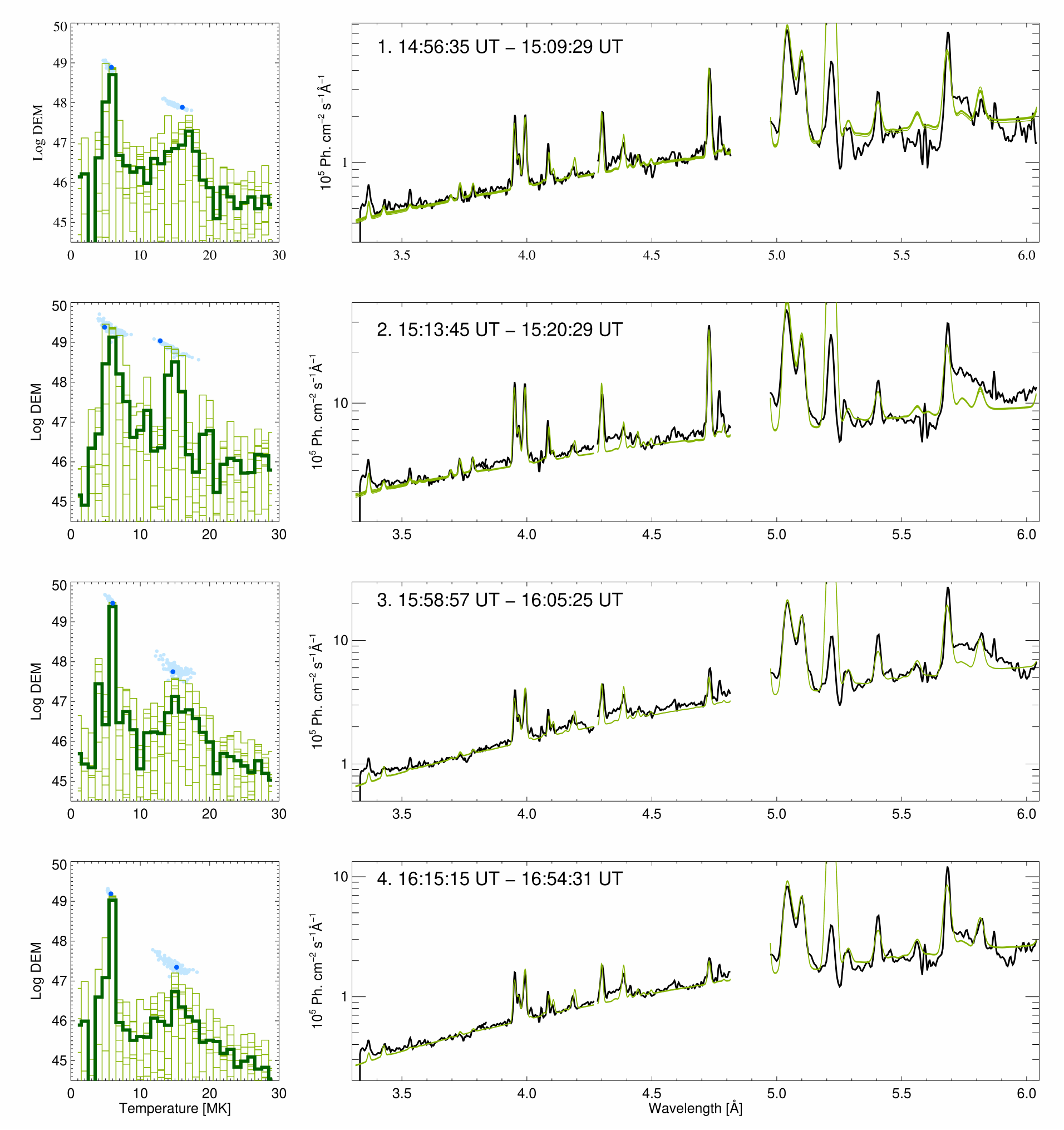}}
\vspace{1cm}
\caption{The sequence of derived DEM distributions (left panel) and RESIK measured and calculated spectra (right) taken
at four selected time intervals. For each time interval 10 DEM from 10 independent generations of the  population are shown (left panel, in light green). The average DEM distributions are presented in dark green. In the right panel the RESIK-observed spectra are presented in black, and synthetic spectra are presented in green. The blue color represents results obtained from the two-temperature model of plasma, the same as presented in Figure 9.}
\label{fig:fig4}
\end{figure}

For further analysis we selected spectra in four time intervals. They are marked as numbers at the top of Figure~9 (the detailed timing is given in the successive Figures 10 and 11). For each of the selected time intervals, we calculated the mean RESIK spectrum and fluxes in the set of 25 wavelength ranges (see Table~2). These fluxes were the input data used for the reconstruction of the DEM distribution and elemental abundances determination. DEM distribution and elemental abundances were calculated simultaneously using the DE method.

At the beginning we used the isothermal model of plasma (values of temperatures, emission measures, and abundances for silicon, sulfur, potassium, and argon) that reproduces the best RESIK fluxes in 25 spectral bands (Table~2)  in each of four chosen time intervals. The results are plotted in the left panel of Figure 10 (in red). Based on these values, the synthetic spectrum was calculated. The comparison of our best-fit calculated spectrum with observations (in black) is given in the right panel. The same procedure has been adopted assuming two-temperature model. The blue color is related to  the two-temperature model of plasma.  The results of calculations  made based on unperturbed observed fluxes are given as dark-red and blue dots. We have also checked how the  data errors influence the solutions. We have randomly perturbed fluxes (10\% error) and determined plasma parameters. The light red and blue represent obtained solutions for 100 independent generations.

One can perceive that for the rise phase of the flare  (time intervals 1 and 2) the temperature that  describes the observation in the isothermal model the best is between the two temperatures derived for the two-temperature model. For the observations during the decay phase (time intervals 3 and 4)  the values for the isothermal model of plasma are closer to the values of the cooler component in two-temperature model of the emitting plasma. A scatter of solutions obtained based on perturbed fluxes for isothermal models is smaller than for the two-temperature model. The values of total emission measure determined for two different models are similar, and the small discrepancy is related to the DE method.

 Next, based on the obtained plasma models from the left panels (temperature, emission measures, and abundances), we calculated the synthetic spectra and compared them to observed ones (right panel of Figure 10). The observations are plotted in black, the red and blue colors correspond to the spectra determined from  isothermal and two-temperature models of emitting plasma, respectively. One can see that the spectra calculated based on the two types of models are almost identical. However, two differences are clearly visible. The first is associated with the continuum level in the shortest wavelengths (3.3~\AA~--~3.6~\AA), the second one with a fit of S~{\sc{xvi}} observed at 4.729~\AA~in wavelength band No. 18 (see Table 2).
 In purple (thin line) the synthetic spectra obtained based on the analysis of GOES fluxes are plotted. The determined values of temperature and emission measure are presented in Table~2. One can see that spectra calculated based on isothermal model of plasma from GOES data and coronal abundances by \cite{Feldman_CHianti_1992ApJS...81..387F}  (Figure 10, in gray) do not agree with RESIK-observed spectra. They are plotted for the comparison only.  We focused on the similarities and differences of the spectra calculated for the isothermal and two-temperature models of plasma from RESIK fluxes.

The results presented for the 30-temperatures (30T) model of plasma  (individuals created according to the scheme from Figure 3b) are shown in  Figure 11 . For each time interval 10 DEM distributions from 10 independent generations of the population are shown (left panel, in light green.) The average DEM distributions are presented in dark green. The results of the analysis using the two-temperature model of plasma are shown in blue for comparison. One can see that these results (2 and 30 temperatures) are not in contradiction. Two DEM components obtained  using 30T models are in agreement with temperature and emission measure values obtained from the two-temperature models.
 
Based on each of 10 independent DEM distributions revealed by the 30T DE exercise, we calculated the respective set of 10 synthetic spectra. These spectra are shown in the right panel of Figure 11 (thin light-green lines). The shapes are so similar that they converge into the slightly wider light-green line falling atop the observed RESIK spectra (in black). It is clearly seen that spectra calculated in the 30T exercise do not reproduce RESIK observations perfectly. The discrepancy between observed and calculated spectra that is seen in wavelength range from 3.3 to 3.6~\AA~may be due to some unknown instrumental effects. The dissimilarity  in spectral range 5~--~6~\AA~ may be associated with the unaccounted for overlapping contribution from higher orders of crystal reflection. Work on this problem (discrepancies) is in progress.

As noted many times, in the DE analysis presented here the DEM distributions and elemental abundances were
fitted simultaneously. The best-fit characteristics (temperature and emission measure) for single- and two-temperature approaches (unperturbed fluxes) are presented in Table~3, together with values obtained from the analysis of GOES fluxes.
The values presented in the parentheses represent limits obtained from perturbed data sets. In the case of the 30T temperature model we calculate the values of average temperature ($T_{av}$) and total emission measure ($EM_{tot}$) for each of the 10 independent DE runs using following formula: $T_{av}= \frac{\int_T T \varphi(T) dT}{\int_T \varphi(T) dT} $, 
$EM_{tot}= \int_T \varphi(T) dT $. The indications ''$1av,  1tot$'' and ''$2av, 2tot$'' are related to temperature ranges 1-10 MK and 10-30 MK, respectively. 

One can see that the temperature and emission measure values obtained from RESIK and GOES in the isothermal
approximation differ substantially. The most probable reason is that GOES determinations (\textsf{goes.pro}, SSW package) use  the standard
''coronal extended'' set of abundances while we allow for abundances to vary, which directly influences the results, and the values of emission measure the most.

The values of the average temperatures and the total emission measure (30T model) are very similar to those obtained from 2T models. It is not a surprising conclusion and reflects the inner consistency of the DE approach.

\begin{table}[h!]
 %\label{tab:table2}
 \begin{center}
\footnotesize
 \caption{The Values of Temperature and Emission Measure Obtained for the 2003 January 21 Flare in Four Selected Time Intervals.}
 \vspace{-0.2cm}
\begin{tabular}{c |c |c |c |c} 
\hline\hline
                   & 1. 14:56:35 UT - 15:09:01 UT & 2. 15:13:45 UT - 15:20:29 UT & 3. 15:58:57 UT - 16:05:25 UT & 4. 16:15:15 UT - 16:42:01 UT\\
          \hline
          & \multicolumn{4}{c}{Temperature [MK]} \\
\hline
GOES &  \textit{T$_{av}$}=11.9 (11.8 -- 13.8) &  \textit{T$_{av}$}=14.3 (13.3 -- 14.8) &  \textit{T$_{av}$}=9.0 (8.9 -- 9.2)    &\textit{T$_{av}$}=8.1 (7.8 -- 8.4) 	\\ 
RESIK 1T      & \textit{T}=8.8 (8.6 -- 9.3)&\textit{T}=10.6 (9.8 -- 11.0) &\textit{T}=6.5 (6.3 -- 6.7)	& \textit{T}=6.3    (6.1 -- 6.6) \\
                
RESIK 2T  &  \textit{T$_{1}$}=5.9 (4.7 -- 6.4)  & \textit{T$_{1}$}=4.9  (3.9 -- 8.6)  &   \textit{T$_{1}$}=6.0  (4.9 -- 6.2)     & \textit{T$_{1}$}=5.8 (5.3 -- 6.3)  \\
                &  \textit{T$_{2}$}=16.1 (13.2 -- 17.4) & \textit{T$_{2}$}=12.9  (11.9 -- 18.4)   &   \textit{T$_{2}$}=14.7  (12.2 -- 17.8)   & \textit{T$_{2}$}=15.2 (11.8 -- 18.2)   \\
RESIK 30T     & \textit{T$_{1av}$}=4.9 (3.2 -- 6.0)  &\textit{T$_{1av}$}=5.7 (4.9 -- 6.1)   & \textit{T$_{1av}$}=5.8 (5.3 -- 6.0)       &\textit{T$_{1av}$}=5.5 (4.3 -- 6.0)  \\
                &\textit{T$_{2av}$}=16.4 (14.4 -- 17.3)  &\textit{T$_{2av}$}=14.9 (14.1 -- 15.4)   & \textit{T$_{2av}$}=16.1 (15.2 -- 17.2)       &\textit{T$_{2av}$}=15.7 (15.0 -- 16.5)  \\
\hline
\hline
          & \multicolumn{4}{c}{Emission measure [10$^{48}$~cm$^{-3}$]} \\
          \hline
GOES &  \textit{EM$_{av}$}=0.74 (0.07 -- 2.26) & \textit{EM$_{av}$}=7.84 (5.2 -- 10.5)&\textit{EM$_{av}$}=6.35 (5.73 -- 6.83)	&\textit{EM$_{av}$}=3.35 (2.57 -- 4.71)	\\ 
RESIK 1T      & \textit{EM}=3.98 (3.62 -- 4.79) & \textit{EM}=18.2 (15.4 -- 20.5)	& \textit{EM}=26.1 (23.7 -- 28.3)	& \textit{EM}= 12.1  (10.4 -- 13.2)    \\
                
RESIK 2T  & \textit{EM$_{1}$}=7.94  (4.88 -- 11.6)   &\textit{EM$_{1}$}=25.1  (12.1 -- 51.0)&  \textit{EM$_{1}$}=31.6   (27.7 -- 48.3)    &\textit{EM$_{1}$}=15.8   (12.1 -- 51.0) \\ 
                & \textit{EM$_{2}$}=0.79 (0.64 -- 1.29)    &\textit{EM$_{2}$}=10.0 (11.3 -- 20.8)  &  \textit{EM$_{2}$}=0.50   (0.30 -- 1.83)    &\textit{EM$_{2}$}=0.22  (0.12 -- 0.60) \\
RESIK 30T     & \textit{EM$_{1tot}$}=11.1   (7.1 -- 20.4)    & \textit{EM$_{1tot}$}=25.8   (22.7 -- 38.0) & \textit{EM$_{1tot}$}=31.9   (30.5 -- 35.3)  &\textit{EM$_{1tot}$}=15.9   (13.5 -- 23.2)  \\ 
                 & \textit{EM$_{2tot}$}=0.76   (0.66 -- 11.1)    & \textit{EM$_{2tot}$}= 7.25   (6.06 -- 8.78)  &\textit{EM$_{1tot}$}=0.55   (0.47 -- 0.60) & \textit{EM$_{2tot}$}= 0.19   (1.69 -- 0.20)  \\          
\hline
\end{tabular}
\end{center}
%\caption{The list of analysed flares.}
    \label{intSt}
\end{table}

The best-fit values of abundance for the four elements considered (Si, S, Ar, and K) calculated for isothermal (1T), two-temperature (2T), and 30-temperature (30T) models of plasma are given in Table~4. For comparison  the values of RESIK abundances obtained in previous analysis by \cite{Sylwester_Opt2015ApJ...805...49S} and by  \citep{Feldman_CHianti_1992ApJS...81..387F} are given.

 \begin{table}[h!]
 %\label{tab:table2}
 \begin{center}
 \caption{Elemental Abundances Obtained from Different Models of Plasma for the 2003 January 21 Flare in Four Selected Time Intervals.}
% \vspace{-0.2cm}
\begin{tabular}{c |c c c c} 
\hline\hline
          &A$_{Si}$&A$_{S}$&A$_{K}$&A$_{Ar}$\\
          \hline
       
1. 14:56:35 UT - 15:09:01 UT &7.43	$\pm$0.07 (1T)&6.99	$\pm$0.05 (1T)&5.00	$\pm$0.03 (1T)&6.37	$\pm$0.05 (1T)\\
                            & 7.51	$\pm$0.07 (2T)&7.10	$\pm$0.08 (2T)&5.04	$\pm$0.22 (2T)&6.38	$\pm$0.08 (2T)\\
	                        & 7.50$\pm$0.04 (30T)&7.10	$\pm$0.02 (30T)&5.03	$\pm$0.10 (30T)&6.40	$\pm$0.01 (30T)\\
\hline
2. 15:13:45 UT - 15:20:29 UT &7.29	$\pm$0.09 (1T)&6.90	$\pm$0.05 (1T)&5.00	$\pm$0.00 (1T)&6.37	$\pm$0.05 (1T)\\
                           & 7.40	$\pm$0.15 (2T)&7.03	$\pm$0.09 (2T)&5.10	$\pm$0.29 (2T)&6.38	$\pm$0.06 (2T)\\
	                       & 7.42$\pm$0.05 (30T)&7.05	$\pm$0.02 (30T)&5.06	$\pm$0.16 (30T)&6.38	$\pm$0.01 (30T)\\
\hline
3. 15:58:57 UT - 16:05:25 UT &7.45	$\pm$0.05 (1T)&7.01	$\pm$0.03 (1T)&5.00	$\pm$0.00 (1T)&6.40	$\pm$0.06 (1T)\\
                           & 7.50	$\pm$0.05 (2T)&7.04	$\pm$0.06 (2T)&5.07	$\pm$0.21 (2T)&6.42	$\pm$0.06 (2T)\\
	                       & 7.51$\pm$0.01 (30T)&7.05	$\pm$0.01 (30T)&5.03	$\pm$0.07 (30T)&6.42	$\pm$0.01 (30T)\\
\hline
4. 16:15:15 UT - 16:42:01 UT &7.48	$\pm$0.05 (1T)&7.02	$\pm$0.04 (1T)&5.00	$\pm$0.00 (1T)&6.40	$\pm$0.07 (1T)\\
                           & 7.50	$\pm$0.07 (2T)&7.04	$\pm$0.04 (2T)&5.07	$\pm$0.33 (2T)&6.41	$\pm$0.07 (2T)\\
	                       & 7.51$\pm$0.01 (30T)&7.05	$\pm$0.01 (30T)&5.03	$\pm$0.14 (30T)&6.41	$\pm$0.02 (30T)\\
	                       
\hline
\cite{Sylwester_Opt2015ApJ...805...49S} & 7.53$\pm$0.05   & 6.97$\pm$ 0.03 &5.74$\pm$0.27 &6.35$\pm$0.04\\
\cite{Feldman_CHianti_1992ApJS...81..387F} & 8.10 & 7.27 & 5.67 & 6.58
\end{tabular}
\end{center}
%\caption{The list of analysed flares.}
    \label{intSt}
\end{table}
The graphical representation results given in Table~3 and Table~4  are  shown in Figure~12.
\begin{figure}[ht!]
 \centerline{\includegraphics[width=15cm,clip=]{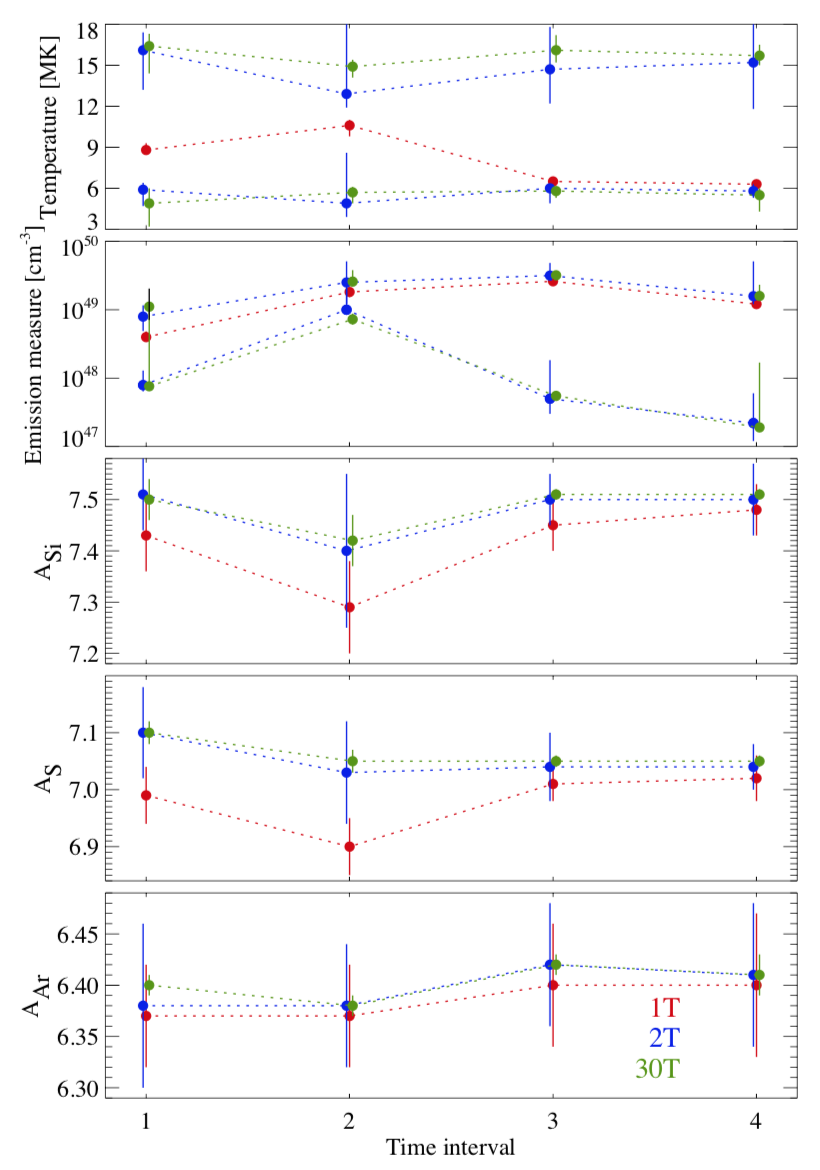}}

\caption{Plasma parameters (temperature, emission measures, and abundances) of the flare on 2003 January 21. The different colors red, blue, and green represent the results obtained  for three different models of plasma:  isothermal, two-temperature, and multithermal (30T), respectively.}
\label{fig:fig4}
\end{figure}
 It is clearly seen that obtained values of abundances are similar and lie within the bounds of errors. They also do not change during flare evolution. The exception is the solution obtained for second time interval~2 (15:13:45~UT~--~15:20:29~UT). Such behavior may be related to nonthermal effects observed during the rise phase of flare. Unfortunately, there are no RHESSI  observation at these times.
The obtained values of elemental abundances for silicon, sulfur, and argon are in agreement  with those obtained by \cite{Sylwester_Opt2015ApJ...805...49S} using WS approach for DEM calculations and the AbuOpt method for determinations of elemental abundances.   However,  potassium abundance could not be determined in our method. This is probably the consequence of the limit of sensitivity of the method and of the before-mentioned problem about disagreement between observed and calculated (based on observed continuum, which is  observed  in  the shortest wavelengths  RESIK channel, (3.3~\AA~--~3.6~\AA). In addition,  the fact that the potassium lines are barely observed only in two spectral bands may cause the problem.

 %The smallest uncertainties were obtained for 30T models of plasma, that is correlated to the most degrees of freedom of solution (the bigger differences in DEM distributions).

\section{Concluding Remarks}
One of the fundamental assumptions when analyzing X-ray spectra  with the aim to calculate DEM distribution  is related to the  elemental abundances. The method discussed in the literature relies on the adoption of the elemental abundance set before calculations of DEM distributions. In such an approach the elemental abundances  errors  translate into inaccuracies in the determined DEM distribution. 

In this research, we introduce the DE approach that resolves this troublesome problem by
simultaneously iteratively improving elemental composition and the DEM shape based on the evolutionary selection process.
The DE algorithm was extensively tested, and the results of the tests confirmed the stability of the solutions and the capability of the algorithm to simultaneously reconstruct the DEM distributions and elemental abundances.

The DE method was applied for the analysis of RESIK spectra obtained for the 2003 January 21 flare. Our aim was to find the plasma model (abundances and distribution of plasma with temperature) that fits RESIK observations the best. We calculated the isothermal model, two-temperature mode,l and DEM distribution described in 30 temperatures (from 1 to 30 MK). 
Our analysis can be summarized as follows:

\begin{itemize}
\item The values of temperature and emission measure calculated based on GOES fluxes using the isothermal model of plasma do not allow us to reproduce RESIK spectra. This (most probably) indicates that coronal set of abundances used by GOES SSW analysis package is not generally applicable for detailed analysis of every coronal source.
\item The two-component model of plasma better describes RESIK observations than the isothermal model (as expected).
\item The values of the average temperatures and the total emission measure (30T model) are very similar to those obtained
for the 2T model.
\item The values of abundance  for silicon, sulfur, and argon do not change during flare evolution.
\item Best-fit DE abundance values are, within uncertainties, the same to those obtained by \cite{Sylwester_Opt2015ApJ...805...49S} for the same flare using a different approach.
\item It was not possible to determine the potassium abundance based on the RESIK spectra used in this study (too small potassium line contribution).
\end{itemize}

We have shown how  the DE algorithm can be used to make breakthrough advance in analysis of
spectra from optically thin plasmas on the example of spectra recorded by RESIK. This approach can be adopted to
analysis of other data sets available from the many instruments observing the Sun (e.g. the Bent Crystal Spectrometer aboard the Solar Maximum Mission  spacecraft) and other stellar objects in the EUV and X-rays in cases in which the source plasma composition is unknown.

When we finished the calculations a new version CHIANTI (10.0.1) became available.
This latest version of CHIANTI includes substantial changes as compared with the previous ones in the line intensities, as well as in the X-ray range. We found these changes for some Si line fluxes to be as large as 10\%. In our future work, we will make use of the most recent version of CHIANTI.
\section{Acknowledgments}
 We acknowledge financial support from the Polish National Science Centre grant No. 2017/25/B/ST9/01821.
 {\software{CHIANTI (v8.0.7; \citep{Dere_1997A&AS..125..149D, Young_2019AAS...23431402Y}), SolarSoft \citep{Freeland_Handy_2012}}} 

\bibliography{bib_sample63_ak}{}
\bibliographystyle{aasjournal}

%% This command is needed to show the entire author+affiliation list when
%% the collaboration and author truncation commands are used.  It has to
%% go at the end of the manuscript.
%\allauthors

%% Include this line if you are using the \added, \replaced, \deleted
%% commands to see a summary list of all changes at the end of the article.
%\listofchanges

\end{document}